\documentclass[aps,prc,twocolumn,superscriptaddress,nofootinbib,tightenlines,
preprintnumbers,showkeys,floatfix]{revtex4-2}

\usepackage[utf8]{inputenc}
\usepackage[english]{babel}
\usepackage{amssymb,amsthm,amsmath,amstext,amsbsy,amsopn,mathrsfs}
\usepackage{bbm}
\usepackage{nicefrac}
\usepackage{slashed}
\usepackage{graphicx}
\usepackage{hyperref}
\usepackage{leftidx}
\usepackage{environ}
\usepackage{mathtools}
\usepackage{xspace}
\usepackage{array}
\usepackage{xcolor}
\usepackage{pgfplotstable}
\usepackage{isotope}
\usepackage{suffix}
\usepackage{booktabs,colortbl}

\newcommand{\ie}{\textit{i.e.}\xspace}

\newcommand{\apriori}{\textit{a priori}\xspace}
\WithSuffix\newcommand\apriori*{\textit{a-priori}\xspace}

\newcommand{\fm}{\ensuremath{\mathrm{fm}}}

\newcommand{\dd}{\mathrm{d}}

\newcommand{\ii}{\mathrm{i}}

\newcommand{\RR}{\mathbb{R}}
\newcommand{\CC}{\mathbb{C}}

\newcommand{\HH}{\mathcal{H}}

\newcommand{\ket}[1]{|#1\rangle}

\newcommand{\braket}[2]{\langle #1|#2\rangle}

\newcommand{\mbraket}[3]{\langle #1|#2|#3\rangle}

\newcommand*\rvec[1]%
{\ensuremath{\overset{\smash{\raisebox{-1.5pt}{\tiny$\rightarrow$}}}{#1}}}
\newcommand*\lvec[1]%
{\ensuremath{\overset{\smash{\raisebox{-1.5pt}{\tiny$\leftarrow$}}}{#1}}}

\newcommand{\MeV}{\ensuremath{\mathrm{MeV}}}

\NewEnviron{subalign}[1][]{%
\begin{subequations}\begin{align}
  \BODY
\end{align}\label{#1}\end{subequations}
}

\NewEnviron{spliteq}{%
\begin{equation}\begin{split}
  \BODY
\end{split}\end{equation}
}

\NewEnviron{mleq}{%
\begin{equation}
  \BODY
\end{equation}
}

\newcommand{\Dil}[2]{D_{#1,#2}}

\newcommand{\maxplus}{\mathbin{\stackrel{\text{max}}{+}}}

\newcommand{\inner}[2]{\langle#1,#2\rangle}
\newcommand{\maxinner}[2]{\inner{#1}{#2}_{\text{max}}}

\newcommand{\dvrsum}[2]{\sum\limits_{#1={-}#2/2}^{#2/2-1}}

\newcommand{\cdd}{{\cdot\cdot}}

\begin{document}

\title{Volume extrapolation via eigenvector continuation}

\author{Nuwan Yapa}
\email{ysyapa@ncsu.edu}
\affiliation{Department of Physics, North Carolina State University,
Raleigh, North Carolina 27695, USA}

\author{Sebastian König}
\email{skoenig@ncsu.edu}
\affiliation{Department of Physics, North Carolina State University,
Raleigh, North Carolina 27695, USA}

\begin{abstract}
 We develop an extension of eigenvector continuation (EC) that makes it possible
 to extrapolate simulations of quantum systems in finite periodic boxes across
 large ranges of box sizes.
 The formal justification for this approach, which we call finite-volume
 eigenvector continuation (FVEC), is provided by matching periodic functions at
 different box sizes.
 As concrete FVEC implementation we use a discrete variable representation based
 on plane-wave states and present several applications calculated within this
 framework.
\end{abstract}

\maketitle

\section{Introduction}

Simulations of quantum systems in finite volume (FV), such as a cubic box with
periodic boundary conditions, can be used to obtain information about that same
system in infinite volume.
In a series of highly influential
papers~\cite{Luscher:1985dn,Luscher:1986pf,Luscher:1990ux}, Lüscher has shown
that the real-world (infinite-volume) properties of the system are encoded in
how its (discrete) energy levels change as the size of the volume is varied.
Bound-state relations connect the finite-volume energy correction
to the asymptotic properties of wave functions, leading to an
exponential volume dependence~\cite{Luscher:1985dn,Konig:2011nz,Konig:2011ti,%
Konig:2017krd}, while information about elastic scattering can be obtained from
discrete energy levels with power-law dependence on the box size.
Resonances, \ie, short-lived, unstable states, are manifest in the
volume-dependent spectrum as avoided crossings of energy
levels~\cite{Wiese:1988qy,Luscher:1991cf,Rummukainen:1995vs}.
While early studies of finite-volume relations considered two-body
applications, work in recent years has focused largely on deriving rigorous FV
quantization conditions for three-body
systems~\cite{Polejaeva:2012ut,Briceno:2012rv,Hansen:2015zga,Hammer:2017uqm,%
Hammer:2017kms,Mai:2017bge,Doring:2018xxx,Pang:2019dfe,Culver:2019vvu,%
Briceno:2019muc,Romero-Lopez:2019qrt,Hansen:2020zhy,Muller:2021uur}, following
early studies of the triton and Efimov trimers in finite
volume~\cite{Kreuzer:2010ti,Kreuzer:2012sr,Kreuzer:2013oya,Meissner:2014dea}.
Related work has derived the volume dependence for bound states comprised of an
arbitrary number of particles~\cite{Konig:2017krd}, and it has been demonstrated
that genuine few-body resonances can be identified from FV
calculations~\cite{Klos:2018sen}, thus providing a discovery tool for such
exotic states.

Eigenvector continuation (EC), first introduced in Ref.~\cite{Frame:2017fah}, is
a powerful (yet strikingly simple in practice) method to address otherwise
unfeasible physics problems.
Given a Hamiltonian with parametric dependence $H(c)$, EC enables robust
extrapolations to a given target point $c_*$ from ``training data'' far away
from that point by exploiting information contained in eigenvectors.
The essence of the system is ``learned'' through the construction of a highly
effective (nonorthogonal) basis, leading to a variational calculation of the
states of interest with rapid convergence~\cite{Sarkar:2020mad}.
Recent work~\cite{Bonilla:2022rph,Melendez:2022kid} has shown that EC as a
particular reduced-basis (RB) method falls within a larger class of model-order
reduction (MOR) techniques.
In practice, EC boils down to constructing Hamiltonian and norm matrices
(denoted as $H(c_*)$ and $N$, respectively) and solving the generalized
eigenvalue problem $H(c_*)\ket{\psi} = \lambda N\ket{\psi}$.

Since its inception, various interesting applications and extensions of EC have
been identified in a short time.
Early applications focusing on bound states include the construction of highly
efficient emulators for uncertainty
quantification~\cite{Konig:2019adq,Ekstrom:2019lss,Sarkar:2021fpz} and robust
extrapolations of perturbation
theory~\cite{Demol:2019yjt,Demol:2020mzd,Franzke:2021ofs}.
More recently, the approach has been extended to construct emulators for
scattering systems~\cite{Furnstahl:2020abp,Melendez:2021lyq,Zhang:2021jmi} and
to studies of nuclear reactions~\cite{Drischler:2021qoy,Bai:2021xok}.

We introduce here a novel extension of EC that goes beyond simple parametric
dependencies of the Hamiltonian.
Specifically, we develop EC as a tool for performing volume extrapolations at
greatly reduced numerical cost.
Since this extension is applicable in connection with any numerical method that
provides access to wave functions in periodic finite topologies, it immediately
yields several interesting applications, among which we highlight in particular
FV studies of few-body resonances~\cite{Klos:2018sen,Dietz:2021haj}.
Identifying such unstable states as avoided crossing of FV energy levels
requires the calculation of spectra over a range of volumes, and in particular in
very large boxes to reach, for example, the low-energy regime of few-neutron
systems, which are of great current interest in nuclear
experiments~\cite{Kisamori:2016jie,Faestermann:2022meh} and nuclear theory (see,
for example,
Refs.~\cite{Gandolfi:2016bth,Higgins:2020pbe,Ishikawa:2020bcs,Dietz:2021haj}).
The technique introduced in this paper provides a way to greatly extend the
reach of FV resonance studies.
Moreover, few-body approaches used to extrapolate Lattice QCD results to
infinite volume via matching to an effective field theory description, recently
discussed in Ref.~\cite{Detmold:2021oro}, can benefit from EC based volume
extrapolation.

\section{Finite-volume eigenvector continuation}
\label{sec:MixedPeriodic}

By ``finite-volume eigenvector continuation (FVEC)'' we refer to the application
of EC to extrapolate properties of quantum states calculated in a set of
periodic boxes with sizes $L_i$, $i=1,\cdd N$ to a target volume $L_*$.
This should be distinguished from using standard EC
at a fixed single volume $L$ to extrapolate a parametric dependence
of the Hamiltonian.
Specifically, we want to consider states $\ket{\psi_{L_i}}$ at volume $L_i$ (or
sets of states $\{\ket{\psi_{L_i}^{(j)}},\,j=1,\cdd N_i\}$) and
perform EC using Hamiltonian and norm matrices
\begin{subalign}[eq:H-N-naive]
 H_{ij} &= \mbraket{\psi_{L_i}}{H_{L_*}}{\psi_{L_j}} \,, \\
 N_{ij} &= \braket{\psi_{L_i}}{\psi_{L_j}} \,.
\end{subalign}
However, at face value the above definitions appear problematic because
the dependence on $L$ does not simply stem from the Hamiltonian; it is inherent
in the definition of the Hilbert space.
Two states $\ket{\psi_{L_i}}$ and $\ket{\psi_{L_j}}$ are actually vectors in
different Hilbert spaces for $i\neq j$, and it is not immediately clear how
the matrix elements written down naively in Eqs.~\eqref{eq:H-N-naive} can be
well-defined quantities.
To resolve this issue, we develop the notion of a vector space that
accommodates states with arbitrary periodicities and show how it relates to
FVEC calculations.

\subsection{Periodic matching}

Let $\HH_L$ be the space of periodic functions $f: \RR \to \CC$ with $f(x+L) =
f(x)$ for some fixed but arbitrary $L>0$.
Consider the union
\begin{equation}
 \HH = \bigcup_{\{L>0\}} \HH_L \,.
\end{equation}
We proceed to show that this concept can be used to define overlaps and matrix
elements of periodic states with different periods.
We restrict the discussion to the special case of a one-dimensional (1D)
two-body system (described by a single relative coordinate $x$), and merely note
that everything generalizes to a larger number of spatial dimensions and/or
particles in a straightforward manner.

\paragraph{Addition.}
Clearly $\HH$ is not a vector space if one defines the sum of $f,g\in\HH$ in the
usual pointwise manner (because the sum of two periodic functions is not in
general periodic).
However, for given $L,L'>0$ one can map $f\in\HH_L$ to $\HH_{L'}$ by means of
a \emph{dilatation}:
\begin{equation}
 (\Dil{L}{L'}f)(x) = \sqrt{\frac{L}{L'}}\,f\!\left(\frac{L}{L'}x\right) \,.
\end{equation}
With this, we can define an addition operation for $f\in\HH_L$ and
$g\in\HH_{L'}$ as follows:
\begin{equation}
 (f \maxplus g)(x) = (\Dil{L}{L'}f)(x) + g(x)
\end{equation}
for $L'>L$, and adjusting $g$ instead in the opposite case.
The result is a periodic function in $\HH_{L'}\subset\HH$, and since
multiplication by a scalar is trivially defined, $(\HH,{\maxplus})$ is a vector
space.

\paragraph{Inner products.}
An inner product on $\HH$ can be defined similarly.
Let $f,g\in\HH$ and, without loss of generality, assume $L \leq L'$ for the
periods of $f$ and $g$, respectively.
Then
\begin{equation}
 \maxinner{f}{g} = \inner{\Dil{L}{L'}f}{g}_{\HH_{L'}}
 = \int_{{-}L'/2}^{L'/2} {(\Dil{L}{L'}f)(x)}^* g(x) \, \dd x
\label{eq:maxinner}
\end{equation}
defines an inner product on $(\HH,\maxplus)$.
Indeed, consider for example adding $h\in\HH_{L''}$ with $L''\geq L'$ to the
second operand:
\begin{spliteq}
 \maxinner{f}{g &\maxplus h} = \maxinner{f}{\Dil{L'}{L''}g + h} \\
 &= \inner{\Dil{L}{L''}f}{\Dil{L'}{L''}g + h}_{\HH_{L''}} \\
 &= \inner{\Dil{L}{L''}f}{\Dil{L'}{L''}g}_{\HH_{L''}}
    + \inner{\Dil{L}{L''}f}{h}_{\HH_{L''}} \\
 &= \maxinner{f}{g} + \maxinner{f}{h} \,,
\end{spliteq}
where we set $x' = (L'/L'')x$ to find
\begin{spliteq}
 \inner{&\Dil{L}{L''}f}{\Dil{L'}{L''}g}_{\HH_{L''}} \\
 &= \int_{{-}L''/2}^{L''/2} \sqrt{\frac{L}{L''}}\,
  f\!\left(\frac{L}{L''}x\right)^{\!*}
  \sqrt{\frac{L'}{L''}} \, g\!\left(\frac{L'}{L''}x\right) \,\dd x \\
 &= \int_{{-}L'/2}^{L'/2} \sqrt{\frac{L}{L'}}\,
  f\!\left(\frac{L}{L'}x'\right)^{\!*}
  g(x') \, \dd x'
 = \maxinner{f}{g} \,.
\end{spliteq}
They key step above was using the property
$\Dil{L}{L''}f = \Dil{L}{L'}\Dil{L'}{L''}f$ of dilatations (which actually form
a multiplicative group).
Other combinations of operands and periods work similarly, and again including
scalar factors is trivial.

\paragraph{Matrix elements.}
Finally, consider a (linear) operator $O$ on $\HH_L$.
While initially this is only given as a mapping $\HH_L \to \HH_L$, we can define
its action on a function $f\in\HH_{L'}$ by inserting an appropriate dilatation:
\begin{equation}
 O f \equiv O \Dil{L'}{L} f \in \HH_L \,.
\end{equation}
Together with the inner product~\eqref{eq:maxinner} this provides a definition
of operator matrix elements between different $\HH_L$, $\HH_{L'}$.

\subsection{Truncated periodic bases}

Consider now truncated bases $S_{L,N}$ and $S_{L',N}$ for $\HH_L$ and
$\HH_{L'}$, respectively, with $N$ a positive integer.
Specifically, let $S_{L,N} = \{ \phi_j^{(L)} : j = 1,\cdd N \}$ with
\begin{equation}
 \phi_j^{(L)}(x) = \frac{1}{\sqrt{L}} \exp\left(\ii\frac{2\pi j}{L} x\right)
\label{eq:PW-basis}
\end{equation}
be a set of plane waves.
Then $\Dil{L}{L'}$ is a bijection between $S_{L,N}$ and $S_{L',N}$, and
because for each $j$ we have $\Dil{L}{L'}\phi_j^{(L)} = \phi_j^{(L')}$.
Therefore, if $\psi$ and $\psi'$ are functions expanded upon $S_{L,N}$
and $S_{L',N}$, respectively, taking the inner product of their coefficient
vectors in $\RR^N$ is the same as considering the inner product on $\HH$ as
defined in Eq.~\eqref{eq:maxinner}.
Note that while this inner product has been defined by matching functions to the
maximum period, we could equally well have chosen to match to the smaller
period.
In practice the concrete choice does not matter because both lead to identical
inner products on $\RR^N$.
Overall we have arrived at a justification for writing down
Eqs.~\eqref{eq:H-N-naive} as well a straightforward prescription for
implementing FVEC numerically.

\paragraph*{Discrete variable representation.}
While conceptually straightforward, the plane-wave
basis~\eqref{eq:PW-basis} is in general not an efficient approach to study
few-body systems.
It can, however, be used as starting point for the construction of a so-called
discrete variable representation (DVR). Originally suggested
as an alternative to harmonic-oscillator based calculations in
nuclear physics~\cite{Bulgac:2013mz}, recent work has established this
plane-wave DVR as a powerful numerical framework for studying few-body
resonances in FV~\cite{Klos:2018sen,Konig:2020lzo,Dietz:2021haj}.
Its construction starts with the states $\phi_j(x)$ defined in
Eq.~\eqref{eq:PW-basis}, with $j={-}N/2,\cdd N/2-1$ for even $N>2$, and where as
before $x$ denotes the relative coordinate for a two-body ($n=2$) system
in $d=1$ dimensions.
Any periodic solution of the 1D Schrödinger equation can be expanded in terms of
the $\phi_j(x)$, yielding a discrete Fourier transform (DFT).
Given a set of equidistant points $x_k \in [{-}L/2,L/2)$ and weights $w_k = L/N$
(independent of $k$), DVR states are constructed as~\cite{Groenenboom:2001aa}
\begin{equation}
 \psi_k(x) = \dvrsum{i}{N} \mathcal{U}^*_{ki} \phi_i(x) \,,
\label{eq:psi-dvr}
\end{equation}
with $\mathcal{U}_{ki} = \sqrt{w_k} \phi_i(x_k)$ defining a unitary matrix.
Calculations in a periodic box can then be carried out through an
expansion in terms of the $\psi_k(x)$ instead of the $\phi_j(x)$.
Importantly, since the transformation between plane-wave states and DVR states
is unitary, the above considerations that justify FVEC carry over to DVR
calculations.

Local potentials are represented in the DVR by basis diagonal
matrices~\cite{Klos:2018sen,Konig:2020lzo}.
Separable potentials have a more complicated representation,
but can also be implemented efficiently~\cite{Dietz:2021haj}.
Another advantage of the DVR is that despite being effectively defined on a
lattice of points, it yields a continuum dispersion relation $E = p^2/(2\mu)$,
where $p$ and $\mu$ are the center-of-mass momentum and the
reduced mass of the system, respectively.
This is achieved by a nondiagonal matrix representation for the kinetic energy
$K$, which is, however, known analytically~\cite{Klos:2018sen,Konig:2020lzo}.
For $d>1$ or $n>2$ the DVR representation of ${K}$
becomes a sparse matrix that can be calculated very efficiently based only on
the 1D two-body matrix elements.
The DVR construction in this case starts from product states of $(n-1)\times d$
plane waves.

As discussed in Refs.~\cite{Klos:2018sen,Konig:2020lzo} it is straightforward
(and numerically very efficient) to construct out of these basic states
subspaces with proper bosonic or fermionic (including spin degrees of freedom)
symmetry properties, and, optionally, definite parity.
Moreover, the breaking of spherical symmetry in infinite volume down to the
cubic symmetry subgroup $O$ in FV can be accounted for by introducing
appropriate projectors~\cite{Johnson:1982yq}, represented as sparse matrices in
the DVR basis~\cite{Klos:2018sen}.
These projectors select a specific cubic irreducible
representation $\Gamma$ out of the set $\{A_1, A_2, E, T_1, T_2\}$ (with
dimensionalities $1$, $1$, $2$, $3$, and $3$, respectively).
Angular-momentum multiplets are reducible with respect to $O$, so
each angular-momentum state in infinite volume in general contributes to
several $\Gamma$.
Low-lying $A_1$ states are to a good approximation dominated by infinite-volume
$S$-wave states, whereas $P$-wave states contribute predominantly to $T_1$
multiplets.
In practice it suffices to perform cubic-projected calculations at selected
volumes in order to assign quantum numbers.

\section{Applications}
\label{sec:Application}

\subsection{Simple two-body system}
\label{sec:2b}

\begin{figure}[tbph] \centering
 \includegraphics[width=0.95\columnwidth]{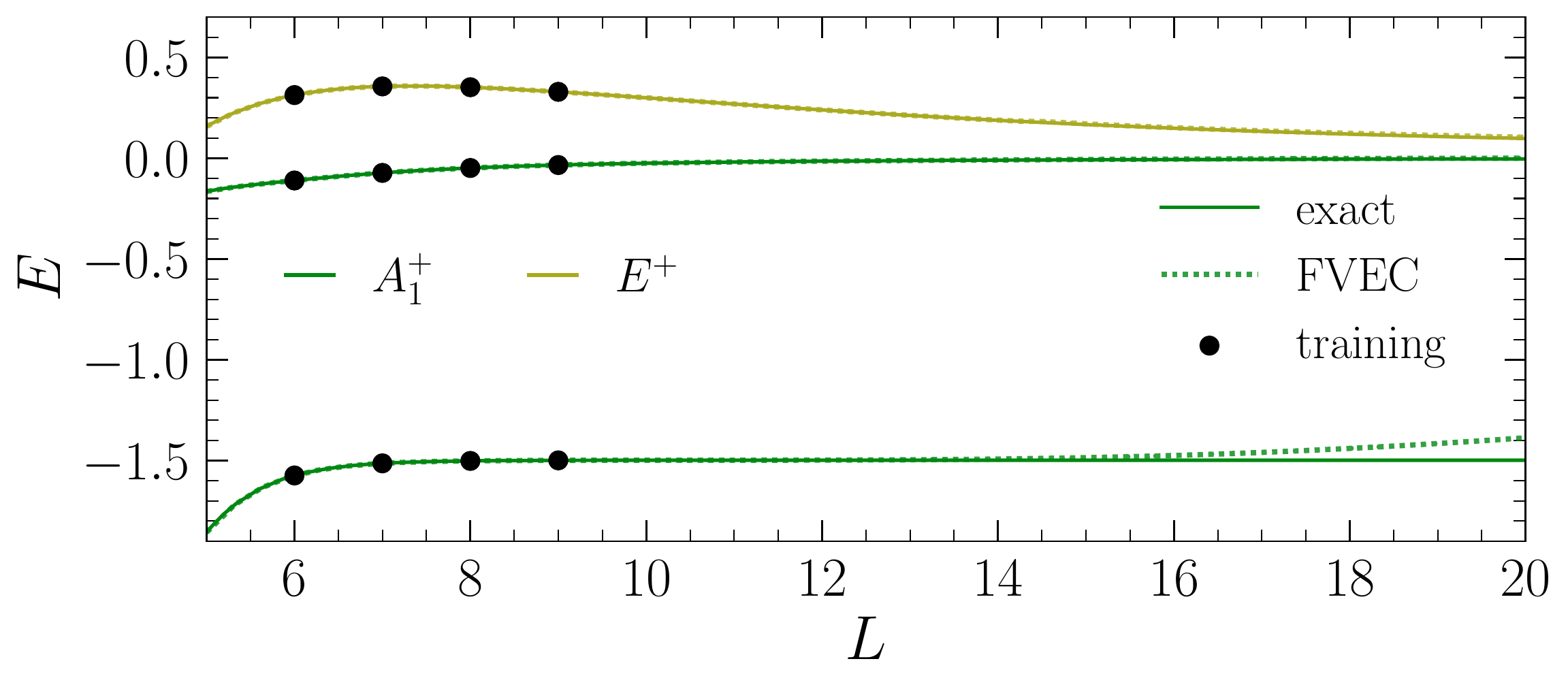}
 \caption{Positive-parity energy spectrum of two particles in finite volume
  as a function of the box size $L$ for a Gaussian potential~\eqref{eq:V-Gauss}
  with $R=2$ and $V_0={-}4.0$ in natural units (see text).  Solid lines show the
  three lowest energy levels calculated in a DVR basis with $N=32$.  Dashed
  lines indicate FVEC results obtained based on training data from four
  different box sizes (solid circles).
  \label{fig:En-d3n2-Pp-EC}
 }
\end{figure}

As a first test we consider a simple two-body system (in three dimensions)
interacting via a Gaussian potential
\begin{equation}
 V(r) = V_0 \exp \biggl(-\Bigl(\frac{r}{R}\Bigr)^{\!2}\biggr) \,.
\label{eq:V-Gauss}
\end{equation}
For this calculation we use natural units with $\hbar = c = 1$ and also set the
particle mass $m=1$.
As (arbitrary) specific choice we set $R=2$ and $V_0={-}4.0$, which produces a
spectrum with two bound $S$-wave states in infinite volume, one of which is very
loosely bound.
In finite volume both bound states are found in the $A_1^+$ representation,
where the superscript indicates positive parity.
The FV spectrum including the lowest states is shown in
Fig.~\ref{fig:En-d3n2-Pp-EC}.
For the FVEC calculation we chose to include training data at four different
volumes, $L=6,7,8,9$, including four states at each training volume so
that the total number of training data is $4\times4=16$.
This covers the two $A_1^+$ bound states as well as the lowest lying scattering
states, falling in the two-fold degenerate $E^+$ representation.
The DVR calculation was performed using an $N=32$ model space for all data
points.
Extrapolation based on this training set work very well, as shown in
Fig.~\ref{fig:En-d3n2-Pp-EC} up to $L=20$, with merely about 4\% deviation
between FVEC and exact calculation for the ground state at $L=20$.

%%%%%%%%%%%%%%%%%%%%%%%%%%%%%%%%%%%%%%%%%%%%%%%%%%%%%%%%%%%%%%%%%%%%%%%%%%%%%%
%%%%%%%%%%%%%%%%%%%%%%%%%%%%%%%%%%%%%%%%%%%%%%%%%%%%%%%%%%%%%%%%%%%%%%%%%%%%%%

\subsection{Three-boson resonance}
\label{sec:3B}

\begin{figure}[tbp] \centering
 \includegraphics[width=0.95\columnwidth]{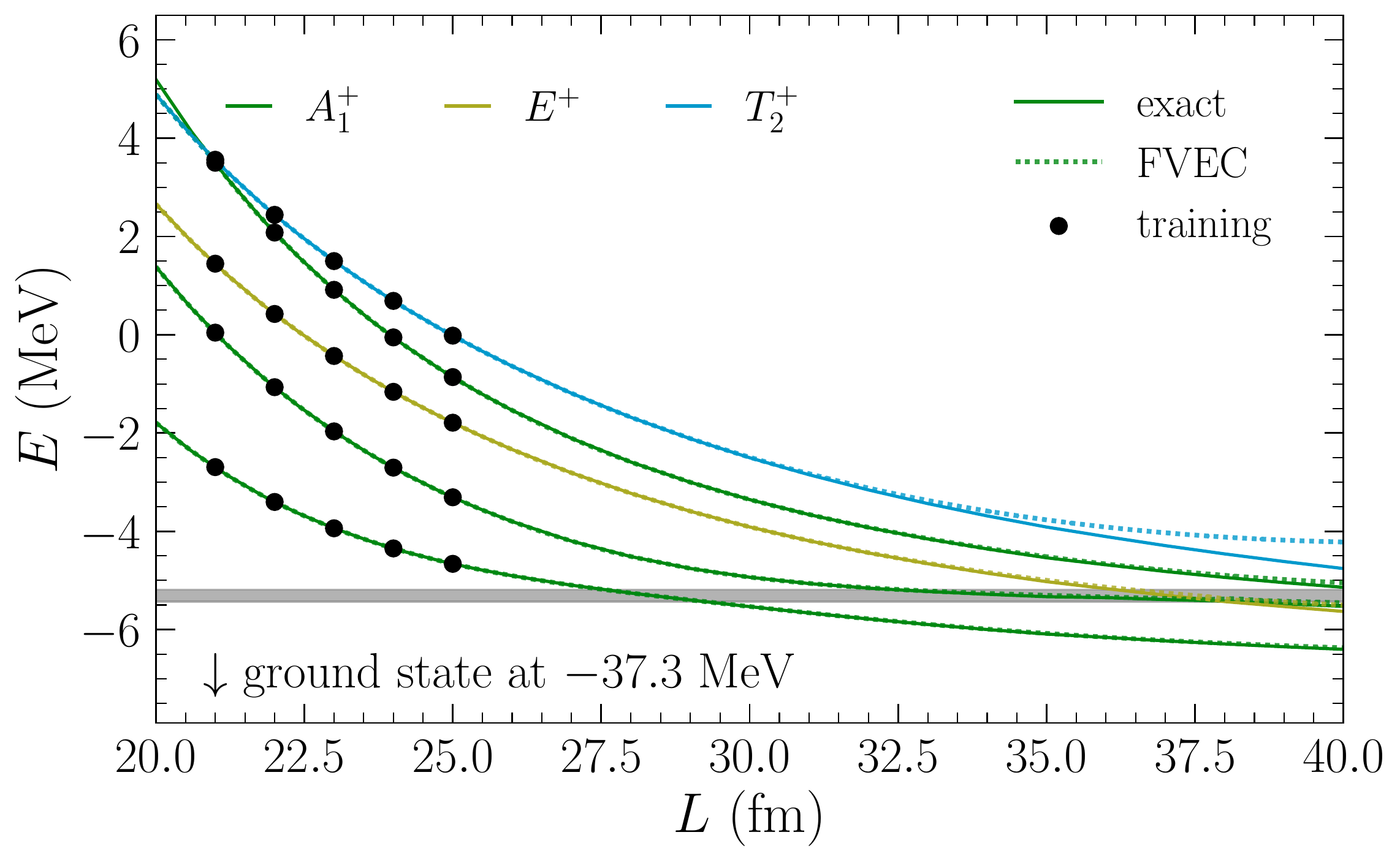}
 \caption{%
  Positive-parity finite-volume energy spectrum of three bosons interacting via
  the potential~\eqref{eq:V-Blandon}.  Solid lines show the exact states
  calculated in DVR bases with $N \leq 28$, whereas dashed lines indicate FVEC
  results obtained based on training data at five different box sizes (solid
  circles).
  The FVEC calculation was
  performed using $8\times5=40$ training states, which includes the
  $A_1^+$ ground state not shown in the plot.  See text for details.
  \label{fig:En-3b-Blandon-1-Pp-EC}
 }
\end{figure}

As another application we consider three identical spin-0 bosons with mass
$m=939.0~\MeV$ (mimicking neutrons) interacting via the two-body potential
\begin{equation}
 V(r) = V_0 \exp\biggl(-\Bigl(\frac{r}{R_0}\Bigr)^2\biggr) + V_1
 \exp\biggl(-\Bigl(\frac{r-a}{R_1}\Bigr)^2\biggr) \,,
 \label{eq:V-Blandon}
\end{equation}
with $V_0=-55~\MeV$, $V_1=1.5~\MeV$, $R_0=\sqrt{5}~\fm$, $R_1=10~\fm$, and
$a=5~\fm$.  This potential produces a resonance state with
energy $E_R=-5.31~\MeV$ and half width $0.12~\MeV$~\cite{Blandon:2007aa} (shaded
band in Fig.~\ref{fig:En-3b-Blandon-1-Pp-EC}).

%%%%%%%%%%%%%%%%%%%%%%%%%%%%%%%%%%%%%%%%%%%%%%%%%%%%%%%%%%%%%%%%%%%%%%%%%%%%%%
%%%%%%%%%%%%%%%%%%%%%%%%%%%%%%%%%%%%%%%%%%%%%%%%%%%%%%%%%%%%%%%%%%%%%%%%%%%%%%

In Fig.~\ref{fig:En-3b-Blandon-1-Pp-EC} we show an FVEC calculation for this
system, using training data at five different box sizes $L=21,22,23,24,25~\fm$
with $N=28$.
For each training volume eight states were included, covering four $A_1^+$
states (including the deeply bound ground state not shown in the figure), one
$E^+$ state, and one $T_2^+$ state (for which only part of cubic multiplet was
included because the training calculations did not all yield the full triplet).
In total, $8\times5=40$ training states were included.
The FVEC calculation provides an excellent reproduction of the exact
energy levels, with noticeable deviations only for excited states at box sizes
far away from the training regime.
In particular, FVEC perfectly captures the avoided crossing between the lowest
two $A_1^+$ states in Fig.~\ref{fig:En-3b-Blandon-1-Pp-EC}, indicating
the three-boson resonance that Ref.~\cite{Klos:2018sen} extracted at
$E_R={-}5.32(1)~\MeV$ from the FV spectrum, in good agreement with
Ref.~\cite{Blandon:2007aa}.

\subsection{Three neutrons}
\label{sec:3n}

Finally, we consider a system of three neutrons ($n$) in pionless effective
field theory at leading order.  Specifically, we use a separable momentum-space
contact interaction,
\begin{equation}
 V(q,q') = C \, g(q) g(q') \,,
 \label{eq:V-sep-q}
\end{equation}
where $g(q) = \exp({-}q^{2n} / \Lambda^{2n})$ is a super-Gaussian regulator.
A projector ensures that the potential acts only on spin-singlet neutron pairs
with vanishing angular momentum (FV analog of the ${}^1S_0$ channel).
This system was recently studied in Ref.~\cite{Dietz:2021haj} (which also
discusses the use of separable interactions with the plane-wave DVR), and as in
that work we set $n=2$ and fix the momentum cutoff $\Lambda = 250~\MeV$.  The
low-energy constant $C$ is fixed to reproduce the $nn$ scattering
length $a_{nn} = {-}18.9~\fm$.

%%%%%%%%%%%%%%%%%%%%%%%%%%%%%%%%%%%%%%%%%%%%%%%%%%%%%%%%%%%%%%%%%%%%%%%%%%%%%%
\begin{figure}[tbp] \centering
 \includegraphics[width=0.95\columnwidth]{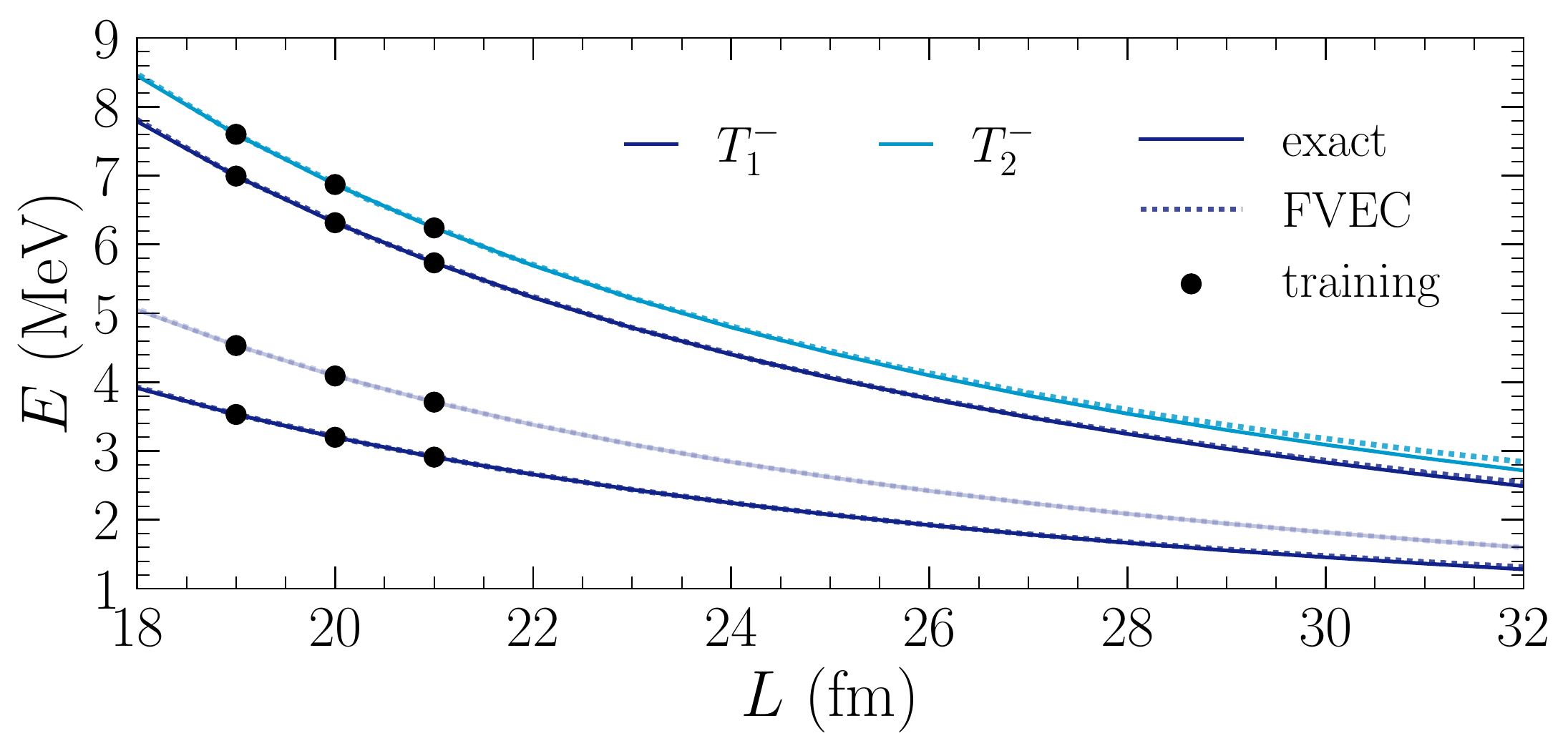}
 \caption{%
  Negative-parity $S_z=1/2$ finite-volume energy spectrum of three neutrons
  interacting via a separable contact potential fit to reproduce the neutron-neutron
  scattering length $a_{nn} = {-}18.9~\fm$.
  Solid lines show the exact states
  calculated in DVR bases with $N \leq 22$, whereas dashed lines indicate FVEC
  results obtained based on $N=22$ training data at three different box sizes
  (solid circles).
  The first and third levels shown in the plots are $T_1^-$ states with total
  spin $S=1/2$.
  The second level is a (noninteracting) $S=3/2$ $T_1^-$ state, whereas the
  fourth level is a $T_2^-$ state with $S=1/2$.
  A total number of $3\times8=24$ training data were used to generate this plot,
  covering a subset of states from the four three-dimensional multiplets (see
  text for details).
  \label{fig:En-3n-SepContact-n4-250-phys-Sz12-Pm-EC}
 }
\end{figure}
%%%%%%%%%%%%%%%%%%%%%%%%%%%%%%%%%%%%%%%%%%%%%%%%%%%%%%%%%%%%%%%%%%%%%%%%%%%%%%

%%%%%%%%%%%%%%%%%%%%%%%%%%%%%%%%%%%%%%%%%%%%%%%%%%%%%%%%%%%%%%%%%%%%%%%%%%%%%%
\begin{figure*}[htbp] \centering
 \includegraphics[width=0.49\textwidth]{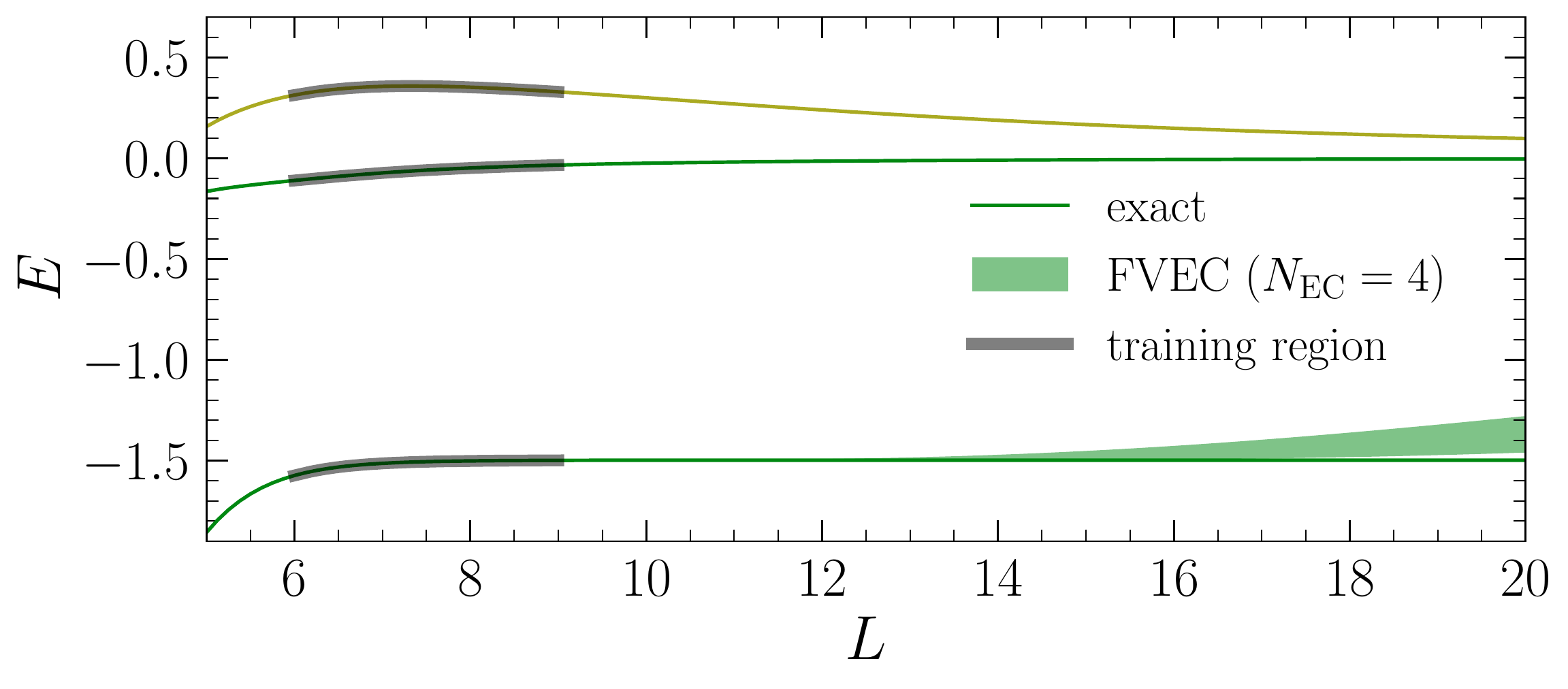}
 \includegraphics[width=0.49\textwidth]{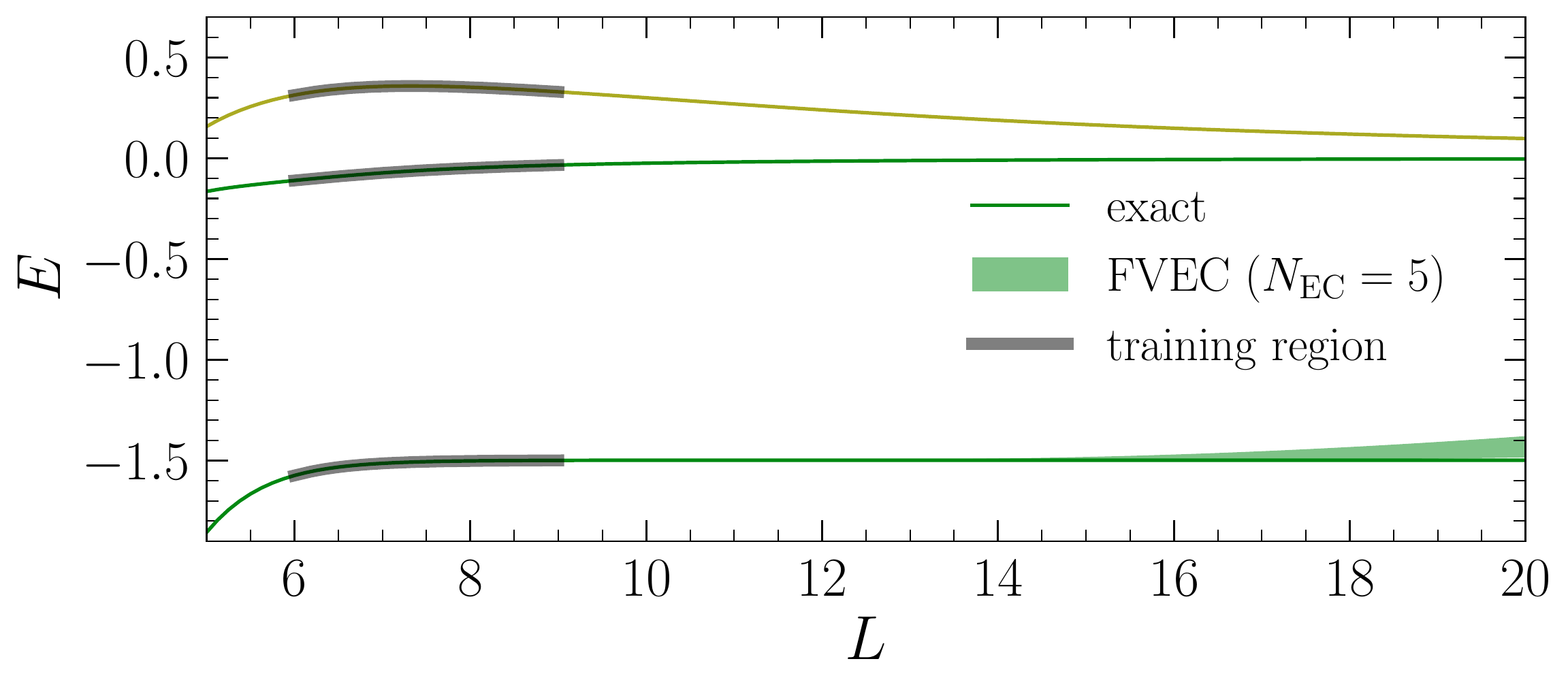}
 \caption{FVEC calculation with uncertainty estimates for two particles
  interacting via a Gaussian potential with range $R=2$ and depth $V_0={-}4.0$
  (in natural units).
  A pool of 16 training data sets with $6 \leq L \leq 9$ (indicated as dark
  shaded bands) was used to estimate the FVEC uncertainty by considering
  all combinations of $N_{\text{EC}}=4$ (left panel) and $N_{\text{EC}}=5$
  (right panel) out of the overall pool.
  The range of all these individual calculations is shown as shaded bands.
  \label{fig:two_body_UQ}
 }
\end{figure*}
%%%%%%%%%%%%%%%%%%%%%%%%%%%%%%%%%%%%%%%%%%%%%%%%%%%%%%%%%%%%%%%%%%%%%%%%%%%%%%

Figure~\ref{fig:En-3n-SepContact-n4-250-phys-Sz12-Pm-EC} shows results
using training data from $N=22$ DVR calculations at $L=19,20,21~\fm$.
The DVR basis $\mathcal{B}$ was restricted to include only states with spin
projection $S_z=1/2$, which covers total spin $S=1/2$ and $S=3/2$.
Its dimension $\dim\mathcal{B} = 28,344,960$ is quite sizable, and even larger
bases are needed to converge the calculation in boxes with
$L\geq32~\fm$~\cite{Dietz:2021haj}.
Compared to the previous examples, this application is more involved because
\textit{(a)} the inclusion of spin increases the DVR basis size at fixed $N$ and
\textit{(b)} the low-lying fermion spectrum is comprised of negative-parity
$T_1$ and $T_2$ states, each coming as threefold degenerate multiplets (with
dominant correspondence to $P$-wave and $D$-wave states in infinite volume,
respectively).
For the training calculations used to generate
Fig.~\ref{fig:En-3n-SepContact-n4-250-phys-Sz12-Pm-EC}, the iterative
diagonalization did not resolve all these degeneracies, finding between one and
three states of each multiplet, not uniform across the different training
volumes.
In spite of these imperfections, FVEC still performs remarkably well after
preprocessing the set of training vectors with a modified Gram-Schmidt
orthogonalization.
This step is well known to be useful for EC calculations in order to avoid
numerical problems stemming from singular and/or ill-conditioned norm matrices.
Therefore, this example demonstrates the robustness of the FVEC method.

\section{Uncertainty estimation}
\label{sec:UQ}

The accuracy of an FVEC calculation depends on the choice of training data,
both on the range it is chosen from and on the number of training points used
to construct the EC subspace.
This dependence can be used to estimate the inherent uncertainty in an FVEC
prediction, which we illustrate in Fig.~\ref{fig:two_body_UQ} for the same
two-body system with attractive Gaussian interaction considered in
Sec.~\ref{sec:2b}.
Instead of using a single fixed set of training points, we calculate (using
$N=32$ for the DVR calculation) a training pool of 16 box sizes located
uniformly within the interval $6 \leq L \leq 9$.
To generate the left panel in Fig.~\ref{fig:two_body_UQ}, we then pick all
possible combinations of $N_{\text{EC}}=4$ training points out of this pool and
perform an FVEC calculation for each of these combinations.
The range of results from these calculations (performed for each target volume)
is shown as shaded bands in Fig.~\ref{fig:two_body_UQ}.
To generate the right panel in the figure the procedure was repeated choosing
all combinations of $N_{\text{EC}}=5$ training points out of the pool of 16.

Accuracy and precision of the extrapolation evidently increase with higher
$N_{\text{EC}}$ as expected.
The band for the ground state almost overlaps at large $L$ with the exact result
for $N_{\text{EC}}=5$, whereas the other levels are already well converged with
$N_{\text{EC}}=4$ (so much so that the shaded bands for the excited states are
barely visible in the figure).
We note that due to the variational nature of EC calculations the bands always
lie above the exact energy levels.
This is a particular feature of energy observables, while no such constraint
holds in general for expectation values of other operators~\cite{Konig:2019adq}.

\section{Discussion and outlook}
\label{sec:Conclusion}

The examples considered above demonstrate that FVEC is able to perform well for
a variety of different scenarios, including bound and unbound states and bosonic
as well as fermionic systems.  In particular, we find the performance of FVEC
roughly independent of the dimension of the model space, considering that all
applications above use comparable numbers of training data.  Based on this one
should expect FVEC to work equally well even at large scales.

Eigenvector continuation has built a reputation of yielding substantial
speedups over exact calculations, to an extent that it can render possible
otherwise unfeasible analyses~\cite{Ekstrom:2019lss}.  FVEC does not disappoint
in this regard: for example, an exact calculation at a single box size shown in
Fig.~\ref{fig:En-3b-Blandon-1-Pp-EC} requires roughly 1100 matrix-vector
multiplications to find the low-energy spectrum of the $N=28$ DVR Hamiltonian
using PARPACK~\cite{PARPACK-ng}.  The FVEC calculation with 40 training data
points on the other hand requires only 40 such matrix-vector products (plus
negligible numerical cost from vector-vector products and solving the EC
eigenvalue problem).  Since the cost of constructing the DVR Hamiltonian for
each target box size is also comparatively negligible, FVEC provides a speedup
factor of roughly 28 for a single $L$ in this particular scenario, and even
more for a calculation spanning multiple $L$ such as shown in
Fig.~\ref{fig:En-3b-Blandon-1-Pp-EC}.

While the focus in the examples we presented has been on using FVEC for
\emph{extra}polation, there is no requirement to choose training data from a
narrow set of volumes.  Sampling instead on both ends of the regime of
interest to perform an interpolation can further improve the accuracy of FVEC
at fixed cost.
Uncertainty estimation as discussed in Sec.~\ref{sec:UQ} works the same
way for this scenario.

Our work provides a perspective for further extensions of EC to scenarios where
the parametric dependence is in the model space rather than just the
Hamiltonian.  In particular, it would be interesting to develop a version of EC
to extrapolate the frequency parameter $\hbar\omega$ in calculations employing
truncated harmonic-oscillator (HO) bases, which play an important role in
nuclear physics.  Such a scheme could for example leverage existing IR and UV
extrapolation schemes~\cite{More:2013rma,Furnstahl:2013vda,Konig:2014hma,%
Furnstahl:2014hca,Wendt:2015nba}.

\begin{acknowledgments}
We thank Dean Lee, Pablo Giuliani, Edgard Bonilla, and Kyle Godbey for useful
discussions and valuable comments on the manuscript.
This work was supported in part by the National Science Foundation under Grant
No. PHY--2044632.
This material is based upon work supported by the U.S. Department of Energy,
Office of Science, Office of Nuclear Physics, under the FRIB Theory Alliance,
Award No. DE-SC0013617.
Computational resources for parts of this work were provided by the Jülich
Supercomputing Center.
Moreover, we acknowledge computing resources provided on Henry2, a
high-performance computing cluster operated by North Carolina State University.
\end{acknowledgments}

\bibliographystyle{apsrev4-1}

\end{document}